\documentclass[aps,prd,floats,floatfix,twocolumn,showpacs,superscriptaddress,nofootinbib]{revtex4}
\usepackage{amsmath, amsfonts}
\usepackage{graphicx}

\begin{document}

\title{Analytical Representation of a Black Hole Puncture Solution}

\author{Thomas W. Baumgarte}

\altaffiliation{Also at Department of Physics, University of Illinois,
Urbana, Il 61801}
\affiliation{Department of Physics and Astronomy, Bowdoin College,
Brunswick, ME 04011, USA}

\author{Stephen G. Naculich}

\affiliation{Department of Physics and Astronomy, Bowdoin College,
Brunswick, ME 04011, USA}

\begin{abstract}
The ``moving puncture'' technique has led to dramatic advancements in
the numerical simulations of binary black holes.  Hannam {\it et.al.}~have 
recently demonstrated that, for suitable gauge conditions commonly employed
in moving puncture simulations, the
evolution of a single black hole leads to a well-known
time-independent, maximal slicing of Schwarzschild.  They
construct the corresponding solution in isotropic coordinates
numerically and demonstrate its usefulness, for example for testing
and calibrating numerical codes that employ moving puncture
techniques.  In this Brief Report we point out that this solution can
also be constructed analytically, making it even more useful as a test case 
for numerical codes.
\end{abstract}

\maketitle

Numerical relativity simulations of binary black holes have recently
achieved a remarkable break-through.  Since Pretorius's 
initial announcement of a successful simulation of binary black hole
coalescence and merger \cite{Pretorius:2005gq}, several groups have
reported similar success
(e.g.~\cite{Campanelli:2005dd,Baker:2005vv,Herrmann:2006ks,Sperhake:2006cy,
Bruegmann:2006at,Scheel:2006gg}).  Of these,
\cite{Pretorius:2005gq,Scheel:2006gg} adopt a ``generalized
harmonic'' formulation of general relativity \cite{Fri85,Gar02} and
eliminate the black hole singularity from the numerical grid with the
help of ``black hole excision''.  All the other groups adopt the BSSN
formulation of the ADM equations \cite{Shibata95,Baumgarte99} together
with a  ``moving puncture'' method to handle singularities.

The idea of the original puncture method was to factor out from the
spatial metric (or more specifically from the conformal factor) an
analytic term that represents the singular terms at a black hole
singularity, and treat only the remaining regular terms numerically.
This approach is very successful for the construction of initial data
(e.g.~\cite{Brandt97}), but did not achieve long-term stable
evolutions in dynamical simulations
(e.g.~\cite{Bruegmann99,Alcubierre00}).  The problem may be associated
with the need for a coordinate system that leaves the puncture -- and
hence the black hole singularity -- at a pre-described location in the
numerical grid, given by the singularity in the analytical function.
The break-through in the recent dynamical puncture simulations is
based on the idea of using a ``moving'' puncture in which no singular
term is factored out.  Care is taken that the singularity never hits a
gridpoint in the numerical grid, but otherwise the puncture is allowed
to move around freely.  With a set of suitable coordinate conditions,
found empirically, this prescription leads to
remarkably stable evolutions.  Clearly, this raises the question how
it can be that the presence of singularities does not spoil the
numerical calculation.  This issue has been clarified recently by
Hannam {\it et.al.}~\cite{Hannam06a,Hannam06b}.

For a single black hole, a moving puncture simulation starts out with
a slice of constant Schwarzschild time expressed in isotropic
coordinates.  These coordinates do not penetrate the black hole
interior, and instead cover two copies of the black hole exterior,
corresponding to two sheets of asymptotically flat ``universes'',
connected by an Einstein-Rosen bridge at the black hole horizon.  The
singularity at isotropic radius $r=0$, where the conformal 
factor $\psi = 1 + M/(2r)$
diverges with $1/r$, corresponds to the asymptotically flat end of the
``other'' universe, and is therefore a coordinate singularity only.

Typically, moving puncture simulations use some variant of the 
``1+log'' slicing condition \cite{Bona97}
\begin{equation} \label{1+log}
(\partial_t - \beta^i \partial_i) \alpha = - 2 \alpha K
\end{equation}
as well as a ``$\tilde \Gamma$-freezing'' condition
\cite{Alcubierre02}.  In \cite{Hannam06a} the authors show that with
these coordinate conditions, the evolution quickly settles down into a
new time-independent solution that is different from the initial data.
In this new time-independent solution the conformal factor features a
$1/\sqrt{r}$ singularity at $r=0$ instead of a $1/r$ singularity,
meaning that the slice has disconnected from the other asymptotically
flat end, and instead terminates on a surface of finite Schwarzschild
coordinate radius.  For the slicing condition (\ref{1+log}),
\cite{Hannam06a} found this limiting surface to be at a Schwarzschild (or areal)
radius of $R = 1.3 M$.  The numerical grid therefore does not include
the spacetime singularity at $R = 0$.  Instead,  
the singularity at $r=0$ is again
only a coordinate singularity, which helps to explain
the success of this numerical method.

In \cite{Hannam06b} (hereafter HHBGSO) the authors consider the
alternative slicing condition
\begin{equation} \label{1+log2}
\partial_t \alpha = - 2 \alpha K.
\end{equation}
Dynamical evolutions again settle down quickly into a new
time-independent solution, which in this case has to be a maximal
slice (with $K = 0$) of Schwarzschild.  A family of time-independent,
maximal slicings of the Schwarzschild spacetime is given by the
following expressions \cite{Estabrook73} for the spatial metric
\begin{subequations} \label{maxslice}
\begin{equation} \label{gamma}
\gamma_{ij} dx^i dx^j = f^{-2} dR^2 + R^2 d^2\Omega,
\end{equation}
the lapse
\begin{equation}
\alpha = f, \label{lapse}
\end{equation}
and the shift vector 
\begin{equation} \label{shift}
\beta^R = \frac{Cf}{R^2}.
\end{equation}
Here we assume spherical polar coordinates, and the function $f$ 
\begin{equation}
f = \left(1 - \frac{2M}{R} + \frac{C^2}{R^4} \right)^{1/2}
\end{equation}
\end{subequations}
depends on both the areal radius $R$ as well as an arbitrary parameter
$C$.  For $C = 0$ we recover Schwarzschild coordinates.
HHBGSO provide evidence that the dynamical evolution, starting from a
slice of constant Schwarzschild time and using the slicing condition
(\ref{1+log2}), settles down to a member of the family
(\ref{maxslice}) with $C = 3 \sqrt{3} M^2/4$, which has a limiting
surface at $R = 3M/2$.  For numerical purposes it is desirable to
obtain this solution in isotropic coordinates.  HHBGSO construct this
solution numerically, and then proceed to show that the solution is
indeed time-independent.  This solution provides a very powerful test
for numerical codes that use the moving puncture method.  Using
this solution as initial data, the time evolution should lead to a 
time-independent solution, so that any deviation from the initial data is 
a measure of the numerical error.

In this Brief Report we point out that this solution in isotropic
coordinates can be constructed analytically, making it an even more
useful tool for code testing.  

To transform the solution (\ref{maxslice}) into isotropic coordinates,
we identify the spatial metric (\ref{gamma}) with its counterpart in
isotropic coordinates,
\begin{equation}
 f^{-2} dR^2 + R^2 d^2\Omega = \psi^4 (dr^2 + r^2  d^2\Omega),
\end{equation}
where $\psi$ is the conformal factor and $r$ the new isotropic radius.
From this identification we have
\begin{equation} \label{psi0}
R^2 = \psi^4 r^2
\end{equation}
and
\begin{equation}
f^{-2} dR^2 = \psi^4 dr^2,
\end{equation}
which we can combine to find
\begin{equation} \label{integral}
\pm \int \frac{dr}{r} = \int \frac{1}{f} \, \frac{dR}{R}
                  =  \int \frac{R\,dR}{\sqrt{ R^4 - 2M R^3 + C^2}}.
\end{equation}
In the general case, the right-hand side may be expressed in 
terms of elliptic integrals of the first and third kinds, 
which of course cannot be written in terms of elementary functions.  
The integral simplifies, however, when two or more roots of the 
quartic equation
$R^4 - 2 M R^3 + C^2 = 0$ 
coincide, which occurs precisely when the discriminant
\begin{equation}
 16\,C^4\left(  16\,C^2  - 27\, M^4\right) 
\end{equation}
vanishes (see, e.g., \cite{Fulton95}).  This happens for two different values of $C^2$.  
In the case $C=0$, integration of (\ref{integral}) 
yields the familiar isotropic form of the Schwarzschild metric.

\begin{figure}[t]
\includegraphics[width=3in]{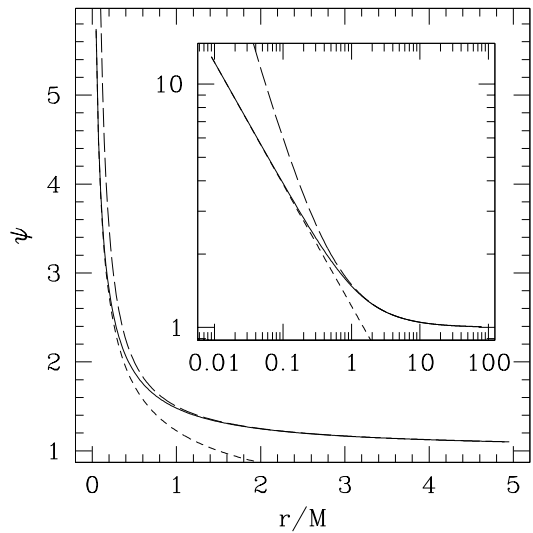} 
\caption{The conformal factor $\psi$ as a function of isotropic radius
$r$ (solid line).  The short-dashed lines shows the asymptotic value
$\psi = \sqrt{3M/(2r)}$ in the limit $r \rightarrow 0$; the
long-dashed line shows the asymptotic value $\psi = 1 + M/(2r)$ in the
limit $r \rightarrow \infty$.  The insert shows the same functions
on a log-log scale.}
\label{fig1}
\end{figure}

Remarkably, the discriminant also vanishes for $C = 3 \sqrt{3} M^2/4$,
the case of particular interest here.
In this case, the quartic polynomial has a double root at $R = 3M/2$,
and the integral can be written 
\begin{equation}
\pm \int \frac{dr}{r} = 
\int \frac{R \, dR}{(R - 3M/2) \sqrt{R^2 + R M + 3 M^2/4}},
\end{equation}
which yields
\begin{eqnarray}
\pm \ln r &=& \frac{1}{\sqrt{2}} 
\ln \left[ \frac{2 R -3 M}
{8 R +  6 M + 3 (8 R^2 + 8 M R + 6 M^2)^{1/2} } \right]  
\nonumber\\
&+ & \ln \left[  2 R + M +  (4 R^2 + 4 M R + 3 M^2)^{1/2} \right] 
+ {\rm const}.
\nonumber\\
&&
\end{eqnarray}
For the positive sign, we then have 
\begin{eqnarray} \label{r}
r & = & \left[ 
\frac{2 R +  M + (4 R^2 + 4 M R + 3 M^2)^{1/2} } {4} 
\right] \nonumber \\
&  \times  &
\left[ \frac{(4 + 3 \sqrt{2})(2 R - 3 M) }
{8 R +  6 M + 3 ( 8 R^2 + 8 M R + 6 M^2 )^{1/2} } 
\right]^{1/\sqrt{2}} \nonumber\\
& = & R \left[ 1 - \frac{M}{R} - \frac{M^2}{2 R^2} + \cdots \right]
\end{eqnarray}
where we fixed the  
constant of integration by requiring that 
$r \rightarrow R$ as $R \rightarrow \infty$.
Eq.~(\ref{r}) gives the value of the isotropic radius $r$ as a
function of the areal radius $R$.  
As expected, we find $r \rightarrow 0$
in the limiting case $R \rightarrow 3M/2$.

\begin{figure}[t]
\includegraphics[width=3in]{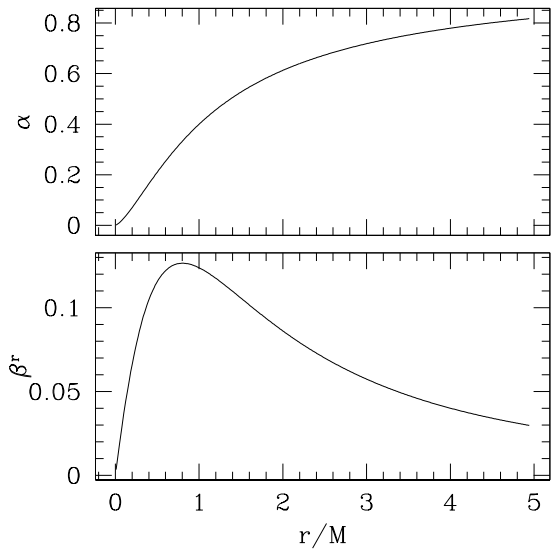}
\caption{The lapse $\alpha$ (top panel) and the shift $\beta^r$
(bottom panel) as a function of isotropic radius $r$ (Compare Fig.~5
in HHBGSO). }
\label{fig2}
\end{figure}

From eqs.~(\ref{psi0}) and (\ref{r}), we obtain the conformal factor 
\begin{eqnarray}
\label{psi1}
\psi 
%&=& \sqrt{\frac{R}{r}} \\find
&=&
\left[ 
\frac
{4 R} 
{2 R +  M + (4 R^2 + 4 M R + 3 M^2)^{1/2} }
\right]^{1/2} \\
&  \times  &
\left[ 
\frac
{8 R +  6 M + 3 ( 8 R^2 + 8 M R + 6 M^2 )^{1/2} } 
{(4 + 3 \sqrt{2})(2 R - 3 M) }
\right]^{1/2\sqrt{2}} \,.
\nonumber
\end{eqnarray}
Although eq.~(\ref{r}) cannot be inverted to yield $R$ as a function of $r$,
eqs.~(\ref{r}) and (\ref{psi1}) 
may be used parametrically to make a plot 
of $\psi$ vs.~$r$  (see Fig.~\ref{fig1}).
Figure~\ref{fig1} also includes the asymptotic limits
\begin{eqnarray}
\psi
&\rightarrow& \left( \frac{3M}{2r} \right)^{1/2} \mbox{~~~~~~~~~}
r \rightarrow 0, \nonumber\\
&\rightarrow& 1 + \frac{M}{2r} \mbox{~~~~~~~~~~~~~~} r \to \infty.
\end{eqnarray}
The lapse 
\begin{equation}
\alpha = \left( 1 - \frac{2M}{R} + \frac{27 M^4}{16R^4}  \right)^{1/2}
\end{equation}
and the isotropic shift
\begin{equation}
\beta^r = \frac{d r}{dR} \, \beta^R = 
\frac{r}{R} \frac{1}{f} \, \frac{C f}{R^2}
= \frac{3 \sqrt{3} M^2}{4} \, \frac{r}{R^3}
\end{equation}
may also be plotted vs.~the isotropic coordinate $r$
(see Fig.~\ref{fig2})
by expressing all quantities parametrically in terms of $R$
(compare Fig.~5 in HHBGSO).

As demonstrated in HHBGSO, this solution for the conformal factor 
$\psi$, the lapse $\alpha$ and the shift $\beta^r$ is time-independent when
evolved with the slicing condition (\ref{1+log2}) for the lapse
$\alpha$ and a $\tilde \Gamma$-freezing condition for the shift
$\beta^i$, which, in this case, leaves the shift constant (see HHBGSO 
for details).  Under these conditions this solution represents the 
asymptotic state of a moving puncture evolution.  This
solution therefore provides a very useful test for moving-puncture
codes: when adopted as initial data, any deviation from these initial
data during the course of the evolution represents a numerical artifact and
hence a measure of the numerical error.
We believe that the analytic form of this
solution presented in this Brief Report will significantly simplify the
implementation and evaluation of such code tests.

\acknowledgments
 
This work was supported in part by 
NSF Grants PHY-0456917 and PHY-0456944 to Bowdoin College.

\end{document}